# Photoluminescence spectroscopy of YVO$_4$:Eu$^{3+}$ nanoparticles with aromatic linker molecules: a precursor to biomedical functionalization

T. R. Senty,[1] M. Yalamanchi,[1] Y. Zhang,[2] S. K. Cushing,[1] M. S. Seehra,[1] X. Shi,[2] and A. D. Bristow[1,a)]

[1] *Department of Physics and Astronomy, West Virginia University, Morgantown, WV 26506-6315, U.S.A.*

[2] *C. Eugene Bennett Department of Chemistry, West Virginia University, Morgantown, WV 26506- 6045, U.S.A.*

Photoluminescence spectra of YVO$_4$:Eu$^{3+}$ nanoparticles are presented, with and without the attachment of organic molecules that are proposed for linking to biomolecules. YVO$_4$:Eu$^{3+}$ nanoparticles with 5% dopant concentration were synthesized via wet chemical synthesis. X-ray diffraction and transmission electron microscopy show the expected wakefieldite structure of tetragonal particles with an average size of 17 nm. Fourier-transform infrared spectroscopy determines that metal-carboxylate coordination is successful in replacing native metal-hydroxyl bonds with three organic linkers, namely benzoic acid, 3-nitro 4-chloro-benzoic acid and 3,4-dihydroxybenzoic acid, in separate treatments. UV-excitation photoluminescence spectra show that the position and intensity of the dominant $^5D_0 - ^7F_2$ electric-dipole transition at 619 nm is unaffected by the benzoic acid and 3-nitro 4-chloro-benzoic acid treatments. Attachment of 3,4-dihydroxybenzoic acid produces an order-of-magnitude quenching in the photoluminescence, due to the presence of high-frequency vibrational modes in the linker. Ratios of the dominant electric- and magnetic-dipole transitions confirm infrared measurements, which indicate that the bulk crystal of the nanoparticle is unchanged by all three treatments.

## I. INTRODUCTION

Some biomedical applications[1] seek to use the luminescence of multifunctional nanostructures for targeted drug delivery, labeling or sensing with magnetic manipulation for separation or capture of cells, proteins and nucleic acids. It is essential that the addition of one function does not hinder another. For example, multifunctional nanoparticles (NP) consisting of magnetic cores and photoluminescent (PL) shells have been demonstrated, using Fe cores coated with rare-earth-doped yttrium vanadate (YVO$_4$) as a luminescent shell.[2]

The advantage of rare-earth-doped crystals is their low toxicity compared to some colloidal quantum dots for biological applications,[3] strong luminescence properties of common ions such as Eu$^{3+}$, Lu$^{3+}$, and Tm$^{3+}$, and background-free emission due to the necessity of indirect excitation via the host material.[4–10] These latter properties are also one of the reasons why rare-earth, particularly lanthanide, doped crystals are used for laser materials and optoelectronics.[11–13] These properties are also promising for use as luminescent nanothermometers.[14,15] YVO$_4$:Eu$^{3+}$ in particular has been shown to have a temperature-dependent linewidth,[16] which can then be exploited to achieve a thermal sensitivity on the nanometer scale. It has been found

---

a) Author to whom correspondence should be addressed. Electronic mail: alan.bristow@mail.wvu.edu

that 5% doping of $Eu^{3+}$ ions in $YVO_4$ leads to strong red PL close to 620 nm.[6]  $YVO_4:Eu^{3+}$ has several other emission lines, including some between 700 – 900 nm[17], which corresponds to the first biological window[18,19]. Because these materials appear ideally suited for nanoscale biomedical applications, it is important that they bind to biomolecules.[20] One possible method of achieving strong binding between the inorganic $YVO_4$ and the organic biomolecules is to use aromatic linker molecules.

In this paper we demonstrate that the native metal-hydroxyl bonds, resulting from the chemical synthesis of the NPs,[21] can be replaced with organic linker molecules and that in some cases the linkers do not affect the desired PL properties. Measurements are performed on as-grown $YVO_4:Eu^{3+}$ NPs, without a magnetic core, to test the validity of the linking and the effect on the PL. We show X-ray diffraction (XRD) and transmission electron microscopy (TEM) structural characterization to confirm the successful creation of the NPs. Fourier-transform infrared (FTIR) spectroscopy confirms attachment of aromatic linkers. PL and photoluminescence excitation (PLE) spectroscopy tests the effect of the treatments on the emission properties.

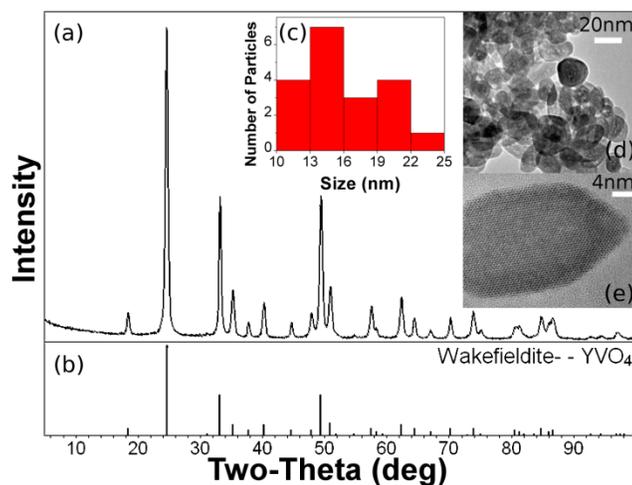

**FIG. 1. (a) X-ray diffraction pattern from the as-grown YVO4:Eu3+ nanoparticles with (b) the expected peak structure given in the International Center for Diffraction Database. Nanoparticle size dispersion histogram (c) measured from diameters. Transmission electron micrographs of samples with (d) 20 nm and (e) 4 nm scale bars.**

## II. Experimental

The sample of $YVO_4:Eu^{3+}$ NPs were created via wet chemical synthesis following the procedures described by Riwotzki and Haase.[4,21] In brief, 3.413g of $Y(NO_3)_3.6H_2O$ and 0.209 g of $Eu(NO_3)_3.6H_2O$ were dissolved in 30 mL of distilled water and stirred for 20 min. This white suspension was placed in an autoclave and was heated to 200 °C. After cooling, the sample was centrifuged for an additional 10 min and the supernatant was discarded. The precipitate was suspended in 40 mL of distilled water and 3.22 g of aqueous 1-hydroxythane-1,1-diphosphonic acid was added. $HNO_3$ was added to this solution to dissolve

Y(OH)$_3$ until a 0.3 pH was reached. The solution was stirred for 1 hour. Concentrated NaOH was added to this suspension until ~12.5 pH was reached, and then it was stirred overnight in a capped vessel. The resulting white suspension was centrifuged for 10 min and the colorless supernatant containing the byproducts was discarded. The process produces a target YVO$_4$ NPs with a Eu$^{3+}$-doping level of 5 at.%, which is known to yield the maximum PL.[6]

For chemical treatments, metal-carboxylate coordination was used to attach the three aromatic linker molecules: benzoic acid (BA), 3-nitro-4-chlorobenzoic acid (NCBA) and 3,4-dihydroxybenzoic acid (DHBA).[22] Specifically, 500 mg of BA, NCBA or DHBA was added to a suspension of 50mg YVO$_4$:Eu$^{3+}$ NPs in 5 mL anhydrous N,N-dimethylformamide and 5 mL anhydrous dichloromethane. The suspension was mechanically shaken at room temperature. After 48 hours, the solid complex was collected by a suction filter and dried by vacuum.

XRD patterns were acquired on powder samples using a Rigaku diffractometer, equipped with CuK$_\alpha$ source at a wavelength of 1.54185 Å and supported by the International Center for Diffraction Database (ICDD) and Jade 9 software. TEM was performed on powder samples using a JEOL JEM-2100 LaB6. For FTIR, PL and PLE spectroscopy measurements, samples were mixed with KBr, at a ratio of sample:KBr = 5:95, and compressed into 0.5-mm-thick disks. FTIR measurements were performed employing an "Infinity Gold" spectrometer by Mattson Instruments in transmission geometry. PL spectra were performed with a double-monochromated Horiba Jobin Yvon Fluorolog-3-11-TCSPC spectrofluorometer in reflection geometry and using an excitation wavelength of 320 nm. PLE measurements were performed using the same instrument, where the excitation wavelength was varied, while monitoring various emission peaks captured in the PL spectra. To measure NP stability, PL spectra were monitored one year apart displaying no difference in intensity or spectral positions. All spectral measurements were taken under ambient conditions.

## III. Results and Discussion
### A. Structural and Surface Properties

Figure 1 (a) shows the XRD patterns captured from the YVO$_4$:Eu$^{3+}$ NPs, revealing Bragg peaks that correspond in angle and magnitude to the wakefieldite or tetragonal zircon crystalline structure expected for bulk YVO$_4$ host; compare to Fig. 1(b) from the ICDD. The Scherrer broadening of the peaks reveals the average crystalline size to be 17 nm and that the structure is single chemical phase. Hence, uniform substitution of Eu$^{3+}$ gives rise to the target of 5 at.% Eu concentration based on the initial weights of the starting materials. Figures 1(d) and (e) show TEM of the as-prepared YVO$_4$:Eu$^{3+}$ crystals with two magnification levels, displaying the expected tetragonal morphology. The distribution of the length of the minor axis of the NP is shown in Figure 1(c), where the average is about 16 nm, obtained using ImageJ software. The minor axis of the NP defines the linewidth in the XRD. Hence, good size agreement is observed between XRD and TEM measurements. From the histogram, a mean aspect ratio was calculated to be 1.58 with the standard deviation of 0.28.

Figure 2 shows the FTIR spectra for the as-grown and treated $YVO_4$:$Eu^{3+}$ NP samples in a KBr matrix. The KBr was measured independently and shows negligible response in the infrared, such that the observed spectral signatures are IR-active modes from the NPs alone. Spectral signatures from the NPs include a peak at 809 $cm^{-1}$ and a shoulder at 920 $cm^{-1}$, which are assigned to the V-O bonds of the $VO_4$ group, and a weaker peak at 452 $cm^{-1}$, which is assigned to the Y-O and Eu-O bonds.[23]

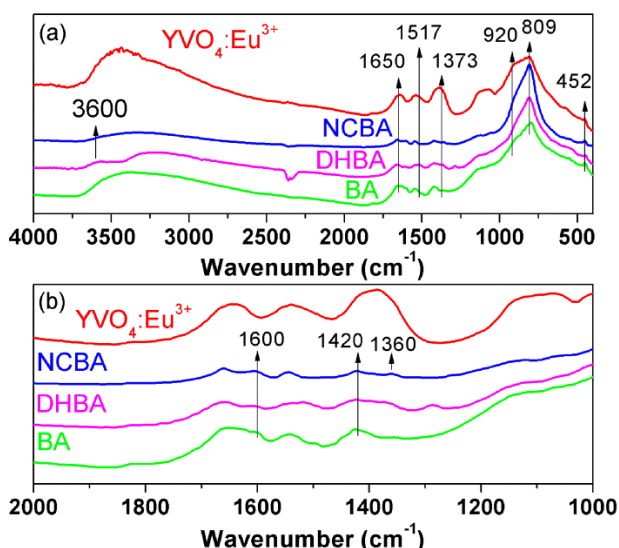

**Fig. 2: Fourier transform infrared spectra of $YVO_4$:$Eu^{3+}$ nanoparticles and $YVO_4$:$Eu^{3+}$ with attached aromatic linker molecules benzoic acid (BA), 3-nitro-4-chlorobenzoic acid (NCBA) and 3,4-dihydroxybenzoic acid (DHBA) for the wavelength ranges (a) 4000 – 400 $cm^{-1}$ and (b) 2000 – 1000 $cm^{-1}$. Spectra are vertically offset for clarity.**

It is important to understand the surface chemical composition for the as-grown NP samples, because this will change when the samples are treated. The as-prepared sample show a broad band feature centered at 3400 $cm^{-1}$ and the sharp peak at 1650 $cm^{-1}$ that are assigned to the $H_2O$ absorbed on to the NP surface.[21] The former band is due to the H-O-H stretch vibrational mode,[24] which is broadened due to the ladder of unresolvable, closely spaced, modes that are modified by the surface interactions. In contrast, the sharp peak is due to the H-O-H bending mode and is unmodified. Some spectral features are due to remnants of the chemical synthesis. The broad shoulder between 1200 $cm^{-1}$ and 1060 $cm^{-1}$ is associated with $POO^-$ impurity from the use of diphosphonic acid.[21] Similarly, the peaks near 1373 $cm^{-1}$ and 1517 $cm^{-1}$ are remaining $NO^{3-}$ surface groups from the use of $HNO_3$.[25]

After all three chemical treatments (NCBA, DHBA and BA) the V-O and Y-O peaks are unaltered, which indicates that in each case there is no modification of the basic crystal structure, as would be hoped when applying a new functional group to a NPs' surface. Figure 2b shows a close up of the IR spectra in the range from 1000 $cm^{-1}$ to 2000 $cm^{-1}$ to assist with the identification of new surface contributions from the treatments. The primary indication of the attachment of linker molecules after the treatments are the new peaks observed at 1600 cm$^{-1}$ and 1420 cm$^{-1}$, which can be assigned to the stretching mode of aromatic ring.[26] Additionally for the NCBA treatment, a peak is observed at 1360 cm$^{-1}$, which can be assigned to N-O bond. A secondary indication that molecules adsorbed on the as-grown NPs are replaced during the treatment is the suppression of the 3400 cm$^{-1}$ and 1650 cm$^{-1}$ features (see Fig. 2a), which correspond to the H-O-H and adsorbed H$_2$O respectively. In particular for the DHBA treatment, the presence of a new band at 3600 cm$^{-1}$ is characteristic of free OH.[27] FTIR results display strong evidence of successful replacement of molecules absorbed on the NP surface and attachment of the three aromatic linker molecules, with no modifications of the YVO$_4$:Eu$^{3+}$ NPs themselves.

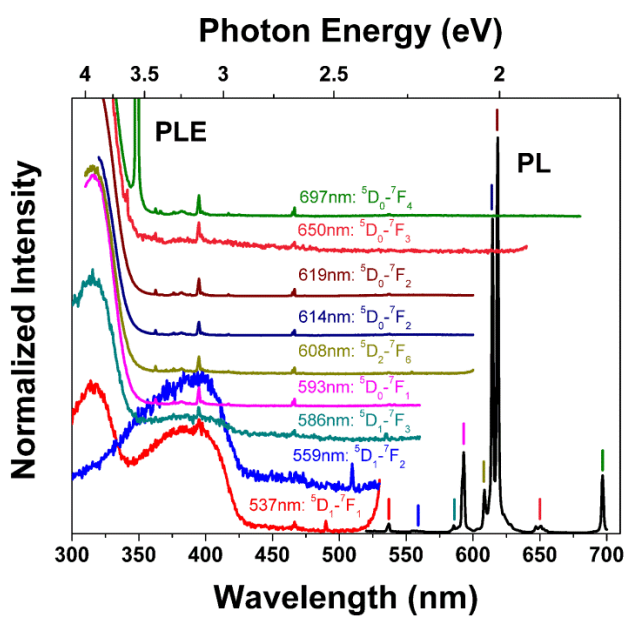

**Figure 3: Photoluminescence (PL) and photoluminescence excitation (PLE) spectra for as-grown YVO$_4$:Eu$^{3+}$.** PL data was captured for an excited wavelength of 250nm and PLE data was monitored at each of the labeled transitions. PLE spectra are vertically offset for clarity.

**B. Optical Properties**

Figure 3 shows the PL and the PLE spectra from the as-grown YVO$_4$:Eu$^{3+}$ NPs. The PL spectrum is acquired with an excitation wavelength of 250 nm, where absorption occurs in the VO$_4^{3-}$ ion.[6] The absorption leads to an excited electron population that can transfer from the vanadate host to excited states of the Eu$^{3+}$ ion by thermally activated migration. Once the excited states of the Eu$^{3+}$ ion are populated, then radiative recombination can occur to provide luminescence. A 5% Eu$^{3+}$ concentration is the tradeoff between the increasing number of luminescence centers and non-radiative concentration quenching that occurs at higher concentrations.[6] As seen in the PL spectrum, emission from the 5% YVO$_4$:Eu$^{3+}$ NPs is strongest between 500 nm and 700 nm. Energy transfer from the host to the Eu$^{3+}$ is confirmed by the lack of emission at the energies corresponding to the states in the $VO_4^{3-}$ ion. For 5% $Eu^{3+}$ the strongest features are near 614 nm and 618 nm, which is a doublet that corresponds to the dominant electric-dipole transition [$^5D_0$-$^7F_2$]. The dominant magnetic-dipole transition [$^5D_0$-$^7F_1$] produces a spectral feature at about 593 nm.[17] The best fit for each peak corresponding to the electric-dipole and magnetic-dipole transitions is provided by a Gaussian lineshape, rather than Lorentzian or Voigt lineshapes. Table 1 shows the extracted spectral center positions and linewidth at full-width half maximum (FWHM). Good agree is found between these values and those quoted in previous literature,[16,17,28,29] and we therefore expect our PL lifetime to be similar too.

Figure 3 also shows PLE spectra recorded for the 697nm, 650nm, 619nm, 614nm, 608nm, 593nm, 586nm, 559nm and 537nm features observed in the PL. Each PLE spectrum is normalized to the strength of the corresponding PL peak, so that the weight of absorption can be compared. The band gap of the $YVO_4$ is approximately 3.5 eV,[30] which corresponds to the sharp rise in the PLE spectra at wavelengths shorter than 350 nm in nearly all the spectra. The presence of this peak confirms that absorption into the vanadate host occurs, prior to the transfer into the $Eu^{3+}$ ions.[5,21] For the PLE spectra corresponding to the PL spectral features at short wavelength, there is an additional lower energy band extending from the band edge out to 425 nm. Absorption occurs in a $[V^{4+}]_A$ defect sites, as has been previously determined by electron paramagnetic resonance spectroscopy.[31] In this case, the energy is sufficiently low to allow for thermal activation of the lowest energy transition in the $^5D_j$ manifold without non-radiative relaxation within that manifold.[32] The 559-nm PL feature is extremely weak, because thermal coupling between the $YVO_4$ and the $Eu^{3+}$ is only allowed in the vicinity of the defect band that has a low density of states.

Figure 4 shows PL spectra for the as-grown and treated $YVO_4$:$Eu^{3+}$ NPs. The general shape of the spectra and spectral center positions, for the NCBA, DHBA and BA treated NPs are unchanged from that of the as-grown NPs, confirming that absorption into and transfer from the vanadate host to the $Eu^{3+}$ ions is still taking place in all treatments. The linewidth at FWHM for the treated NPs are also shown in Table 1 for comparison to the as-grown NPs. Peak positions are unchanged for all treatments and the linewidths at FWHM do not change significantly for the NCBA and BA treatments, but there is a slight trend of a smaller linewidth for the DHBA treatment.

**TABLE I. Spectral Properties of electric-dipole and magnetic dipole transitions.**

| Transition | Line Center | Full-Width at Half Maximum | | | |
|---|---|---|---|---|---|
| | | $YVO_4$:Eu | NCBA | DHBA | BA |
| $^5D_0$-$^7F_2$ | 614.5nm | 34.7cm$^{-1}$ | 33.9cm$^{-1}$ | 30.2cm$^{-1}$ | 33.4cm$^{-1}$ |
| | 618.4nm | 36.6cm$^{-1}$ | 35.5cm$^{-1}$ | 29.9cm$^{-1}$ | 35.5cm$^{-1}$ |
| $^5D_0$-$^7F_1$ | 591.8nm | 26.4cm$^{-1}$ | 24.6cm$^{-1}$ | 23.2cm$^{-1}$ | 25.4cm$^{-1}$ |
| | 593.1nm | 36.3cm$^{-1}$ | 34.5cm$^{-1}$ | 25.4cm$^{-1}$ | 31.5cm$^{-1}$ |

A more detailed investigation of the electromagnetic environment of the $Eu^{3+}$ ion can be determined through the intensities of the $^5D_0$-$^7F_2$ electric-dipole transition and $^5D_0$-$^7F_1$ magnetic-dipole transition. If the $Eu^{3+}$ ion is centrosymmetric in the lattice then the magnetic-dipole transition dominates, whereas any asymmetry in its position will make the electric-dipole mode dominate.[33,34] For all samples, the electric-dipole transition dominates, indicating that the $Eu^{3+}$ ion sits at a non-inversion symmetry site. Moreover, comparing the ratio of the $^5D_0$-$^7F_1$ and $^5D_0$-$^7F_2$ transitions intensities gives a measure of the amount of distortion[35–37] from the as-grown $YVO_4$:$Eu^{3+}$ NPs by the treatments. The intensity ratios are 0.2050, 0.1961, 0.220 and 0.2048 (with error of ± 0.0006) for the as-grown $YVO_4$:$Eu^{3+}$, NCBA-, DHBA- and BA-treated NPs respectively. Hence, no significant distortion is present, as might be expected and agreeing with the FTIR results discussed above. This means that only surface chemistry is affected by the treatments.

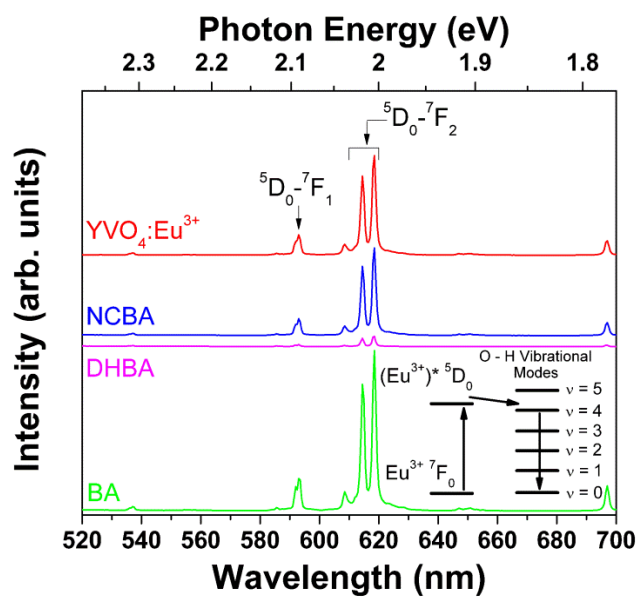

**Fig. 4 Emission spectra of as-grown and treated YVO4:$Eu^{3+}$ nanoparticles excited at 250 nm. The attached aromatic molecules are benzoic acid (BA), 3-nitro-4-chlorobenzoic acid (NCBA) and 3,4-dihydroxybenzoic acid (DHBA). Spectra are vertically offset for clarity. The inset shows a schematic diagram of alignment of $Eu^{3+}$ energy gap with O-H vibrational modes.**

The primary difference between the spectra is the overall emission strength. The spectral density and weights are nearly identical for the treated and as-grown NP samples, although some degree of PL suppression is observed when adding the aromatic linker molecules. This is most notably for the DHBA-treated sample. The spectra have been calibrated to account for differences in molecular surface coverage and NP concentrations. The uniform quenching of the PL is therefore an indication of the addition of a non-radiative pathway from the $Eu^{3+}$ ions. It is known that hydroxyl groups quench luminesce from the $Eu^{3+}$ ions,[38,39] which is why the DHBA-treated sample has the strongest suppression. High-frequency O-H vibrational modes couple with the excited state of the $Eu^{3+}$ ion; see the inset of Fig. 3. The quenching is strongest when the radiative transition of the lanthanide dopant is close in energy to an integer number of O-H vibrational quanta and when the quantum number (ν) is smallest.[40] In the case of $Eu^{3+}$ ions, this integer is only ν = 4 and the energy matching is very close, leading to rapid non-radiative relaxation through the hydroxyl groups. Confirmation of free O-H groups in the DHBA-treated sample is given at 3600 cm$^{-1}$ in the FTIR spectrum, but these modes are not observed in the spectra for the samples with BA and NCBA treatments; see Fig. 2. Overall the BA- and NCBA-treated NP samples do not exhibit the same degree of PL quenching, making them preferable for further bio-functionalization.

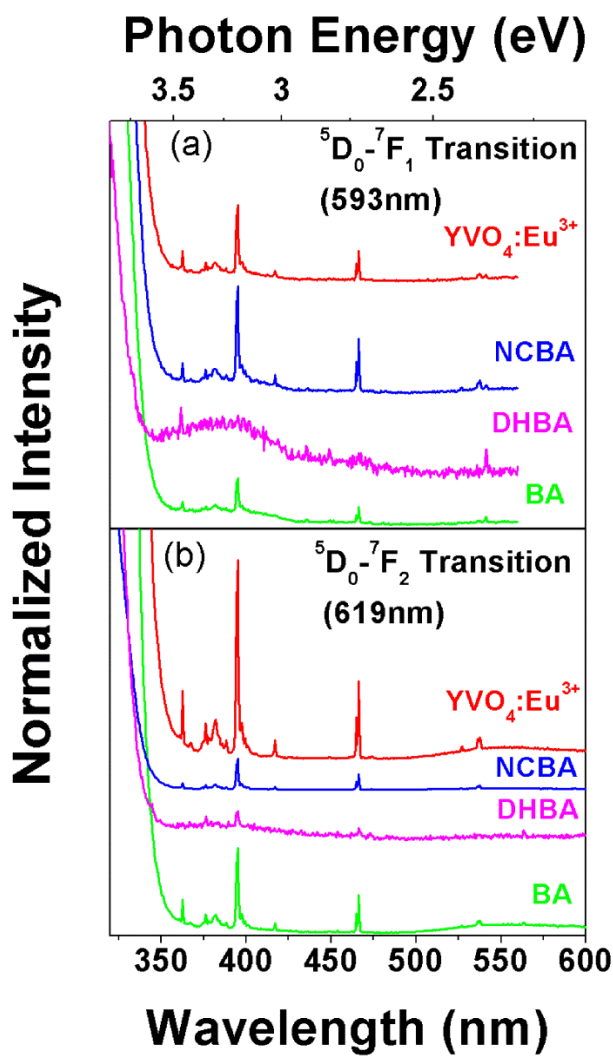

**Fig. 5:** Photoluminescence excitation spectra for as-grown and treated $YVO_4:Eu^{3+}$ nanoparticles for the a) magnetic-dipole $^5D_0$-$^7F_1$ transition at 593nm and b) electric-dipole $^5D_0$-$^7F_2$ transition at 619nm. Spectra are normalized to the peak emission at the respective wavelengths and offset for clarity.

Figure 5 shows the PLE spectra for (a) the magnetic-dipole transition [$^5D_0$-$^7F_1$] and (b) the electric-dipole transition [$^5D_0$-$^7F_2$] for all four NP samples. The data are all normalized to the emission peaks at (a) 593 nm and (b) 619 nm respectively, which provide a comparison of the excitation strengths. Comparing the three treated samples, to the as-grown YVO$_4$:Eu$^{3+}$ NPs, reveals that absorption occurs in all samples at wavelengths shorter than 350 nm into the excited states of the VO$_4^{3-}$ ions. In all cases, this absorption is nearly the same. The fact that this comparison holds for the DHBA-treated NP sample confirms that the actual suppression of the emission is indeed due to relaxation in the O-H vibration ladder and not due to reduced absorption into the vanadate host. For the DHBA-treated sample there is also a slight increase in the relative contribution of the absorption into the defect band, as coupling to the Eu$^{3+}$ ions is less affected by the relaxation pathway introduced by the O-H vibration ladder. The PLE spectra also show direct excitation of the Eu$^{3+}$ ions. This is strongest in the as-grown nanoparticles and somewhat suppressed in the BA- and NCBA-treated NP samples. The strongest suppression is once again for the DHBA-treated NPs, further indicating the suppression of the effectiveness of the Eu$^{3+}$ ions in the presence of the hydroxyl groups.

From these optical and FTIR studies, it is seen that aromatic linker molecules can be added as a precursor to functionalizing YVO$_4$:Eu$^{3+}$ NPs, as long as there are no hydroxyl groups present. Some suggestions include adding biotin and DNA. For functional molecules containing hydroxyl groups, a dependence on the distance between the Eu$^{3+}$ emitters and the source of the O-H vibration ladder will have to be performed.

## VI. Conclusion

In this paper, YVO$_4$:Eu$^{3+}$ nanoparticles have been synthesized and characterized. The crystalline structure and nanoparticles size have been determined. Aromatic linker molecules of benzoic acid, 3-nitro-4-chlorobenzoic acid and 3,4-dihydroxybenzoic acid have been attached to the surface replacing the native surface molecules in the as-grown sample. The attachment is confirmed by Fourier-transform infrared spectroscopy. The aromatic treatments do not change the photo-physics of the nanoparticles, except in the case of the addition of hydroxyl groups by providing a non-radiative relaxation pathway that suppresses the luminescence from the emitting lanthanide ions.

These studies show that these YVO$_4$:Eu$^{3+}$ nanoparticles are suitable for functionalization in nanoscale biomedical applications, where luminescence is important for the application. Moreover, since it has been shown previously that rare-earth doped vanadate crystals can be used to coat magnetic core for magneto-luminescence applications, we suggest that this study is an important step for those applications too.

**ACKNOWLEDGMENTS**

This research was sponsored by an Award for Research Team Scholarship from the Eberly College of Arts and Science at West Virginia University. SKC was supported by NSF Research Graduate Fellowship (#1102689).